\documentclass[letterpaper,aps,prl,twocolumn,english,showpacs,superscriptaddress]{revtex4-1}

\PassOptionsToPackage{
	breaklinks=true,
	colorlinks=true,
	citecolor=	blue,
	linkcolor=	blue,
	urlcolor=	blue}{hyperref}

\usepackage{hyperref} 
\usepackage{amsmath}
\usepackage{amsfonts}
\usepackage{footnote}
\usepackage{todonotes}
\usepackage[caption=false]{subfig}
\usepackage{xcolor}
\usepackage{graphicx}
\usepackage[squaren]{SIunits}
\usepackage{IEEEtrantools}

\PassOptionsToPackage{
	breaklinks=true,
	colorlinks=true,
	citecolor=	blue,
	linkcolor=	blue,
	urlcolor=	blue}{hyperref}

\begin{document}

\title{Atom Interferometry in a Warm Vapor}

\author{G. W. Biedermann}
\email{gbieder@sandia.gov}
\affiliation{Sandia National Laboratories, Albuquerque, New Mexico 87185, USA}
\affiliation{Center for Quantum Information and Control (CQuIC), Department of Physics and Astronomy, University of New Mexico, Albuquerque, New Mexico 87131, USA }
\author{H. J. McGuinness}
\affiliation{Sandia National Laboratories, Albuquerque, New Mexico 87185, USA}
\author{A. V. Rakholia}
\author{Y.-Y. Jau}
\affiliation{Sandia National Laboratories, Albuquerque, New Mexico 87185, USA}
\affiliation{Center for Quantum Information and Control (CQuIC), Department of Physics and Astronomy, University of New Mexico, Albuquerque, New Mexico 87131, USA }
\author{D. R. Wheeler}
\affiliation{Sandia National Laboratories, Albuquerque, New Mexico 87185, USA}
\author{J. D. Sterk}
\affiliation{Sandia National Laboratories, Albuquerque, New Mexico 87185, USA}
\author{G. R. Burns}
\affiliation{Sandia National Laboratories, Albuquerque, New Mexico 87185, USA}

\date{\today}

\begin{abstract}
We demonstrate matterwave interference in a warm vapor of rubidium atoms.  Established approaches to light pulse atom interferometry rely on laser cooling to concentrate a large ensemble of atoms into a velocity class resonant with the atom optical light pulse.  In our experiment, we show that clear interference signals may be obtained without laser cooling.  This effect relies on the Doppler selectivity of the atom interferometer resonance.  This interferometer may be configured to measure accelerations, and we demonstrate that multiple interferometers may be operated simultaneously by addressing multiple velocity classes. 
\end{abstract}

\maketitle

The technique of light pulse atom interferometry (LPAI) has proved to be exceptionally useful for precision acceleration measurements.  Since its inception \cite{Kasevich1991}, research has branched into pursuits of inertial sensor technology \cite{Muller2009, McGuinness2012, Rakholia2014, Bodart2010, Butts2011} and foundational precision measurements \cite{Dimopoulos2007,Schlippert2014, Fixler2007, Rosi2014}, including space-based gravity wave detectors \cite{Graham2013}.   These demonstrations build upon well-vetted techniques in the field of laser cooling and trapping \cite{Metcalf}.  Reducing the velocity distribution of a large ensemble of atoms and collecting them into a well-defined spatial location affords ample time for interrogation \cite{Dickerson2013, Muntinga2013} and high fidelity detection \cite{Biedermann2009}.  In this setting, the matter wave of each atom evolves with inertial freedom such that photon recoils may be used to coherently split and recombine the wave packets without perturbation.  The experimental overhead is laser system complexity and ultra-high vacuum requirements that have challenged efforts fielding these instruments \cite{Wu2009, Stockton2011, geiger2011, Muntinga2013, Lautier2014, Biedermann2015}.  

The simplicity of a vapor cell approach, used for atomic clocks \cite{Camparo2007} and magnetometry \cite{Kawabata2010, Seltzer2009}, is an alluring alternative. In this approach, long interrogation times are achieved through the use of a buffer gas or a spin anti-relaxation coating.  As such, multiple collisions occur between the interrogated atom and the buffer gas or cell coating over the duration of one measurement period.  Such collisions spoil the inertial purity of the wave packets and would obfuscate the LPAI fringe.  Nevertheless, by borrowing certain aspects of the vapor cell approach, namely a spin anti-relaxation coating for state preparation, and blending this with the inherent velocity-filtering function of the photon recoil in LPAI, we re-imagine atom interferometry.  Consequently, we achieve high fidelity interference signals in a significantly simplified warm vapor experiment, without laser cooling.   

\begin{figure}
\includegraphics[width=3.5 in]{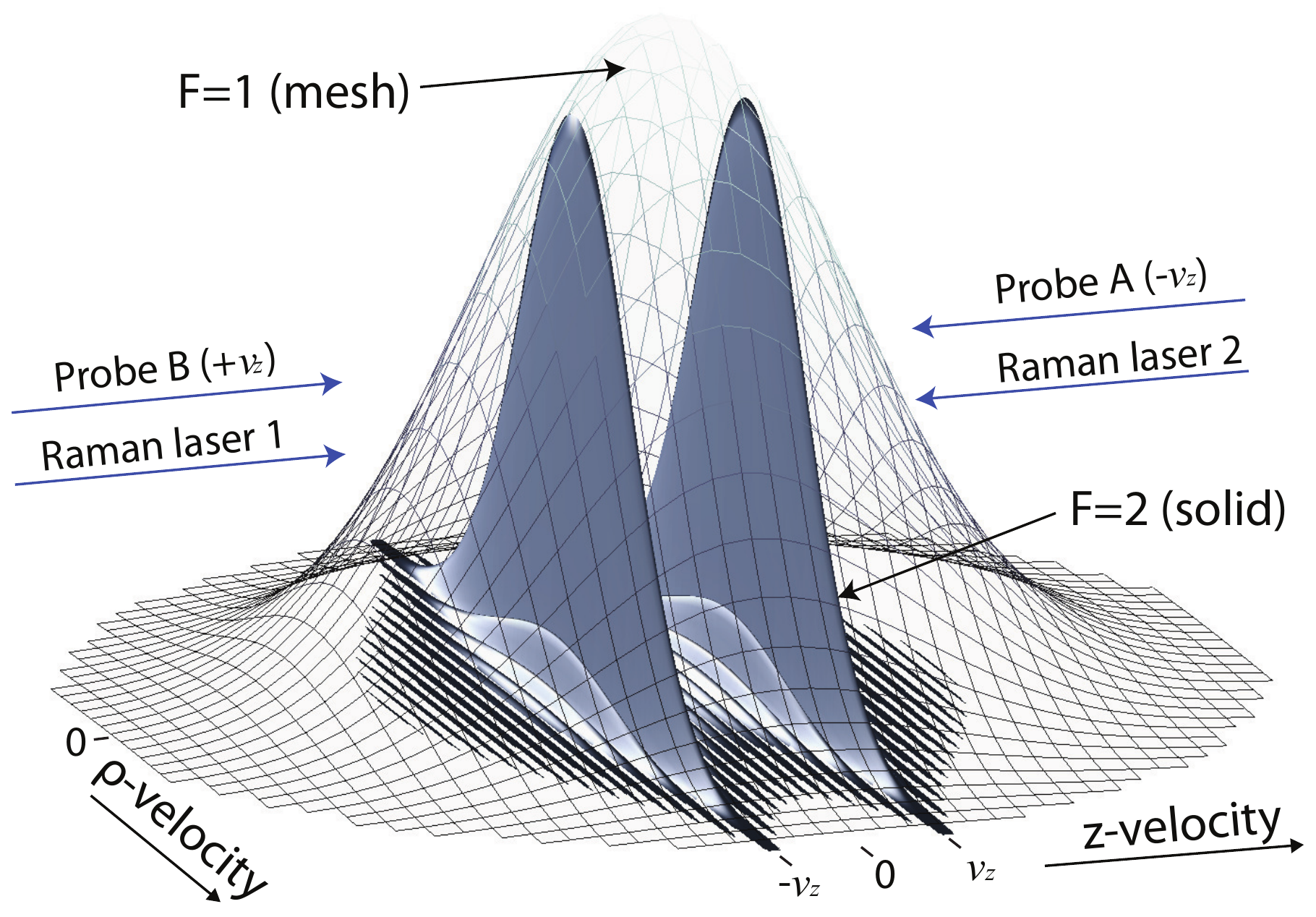}
\caption{\label{fig:fig1} (Color online) Vapor interferometer concept--not to scale. The 2-D mesh Gaussian represents the  Maxwell-Boltzmann distribution in cylindrical coordinates $z$ and $\rho$ for  room temperature atoms in $|\text{F} = 1\rangle$. The solid blue Sinc functions are two narrow velocity classes centered at $\pm v_z$ in $|\text{F} = 2\rangle$ that are selected from $|\text{F} = 1\rangle$ via the Raman transitions that comprise the LPAI.  The arrows indicate the directions along $\hat{z}$ of the Raman and probe lasers used to generate and detect the two narrow classes.  Each Raman laser carries two frequencies separated by the hyperfine splitting, $\nu_{hf}$ plus an additional amount equal to $|\textbf{k}_{\textrm{eff}}| v_z/\pi$ such that a counter-propagating two-photon Raman transition is simultaneously resonant with the two velocity classes. Following the LPAI, the two velocity classes are simultaneously detected with two resonant probe lasers.}
\end{figure}

LPAI uses two-photon stimulated Raman transitions between hyperfine ground states (e.g. $|\text{F} = 1\rangle$ and $|\text{F} = 2\rangle$ in $^{87}$Rb) to create coherent superpositions of momentum states with the effect of redirecting matter wave packets to form the atom optical elements of beam splitter and mirror.  When the two optical fields are arranged in a counter-propagating geometry, the transition has the velocity sensitivity of an optical transition with wavevector, $|\textbf{k}_{\textrm{eff}}| \approx $ 4$\pi/\lambda$ where typically $\lambda$ = 780 nm for $^{87}$Rb \cite{Kasevich1991}.  For a given Rabi frequency, $\Omega_R$, of the Raman transition, a velocity class with a Doppler width of approximately $\Omega_R/\textbf{k}_{\textrm{eff}}$, is filtered from the thermal distribution (see \autoref{fig:fig1}) \cite{Moler1992}.  For atoms in state $|F = 1\rangle$, driving the stimulated Raman transitions in a $\pi/2-T-\pi-T-\pi/2$ sequence forms a Mach-Zehnder atom interferometer where the probability for an atom to be in atomic state $|\text{F} = 2\rangle$ following the pulse sequence is given by $\textrm{P}_{|\textrm{F}=2\rangle} = \frac{1}{2}(1+ \cos(\Delta\phi))$  \cite{Berman1997}.  For an atom undergoing an acceleration $\bf{a}$, it follows that $\Delta\phi = -\textbf{k}_{\textrm{eff}} \cdot \textbf{a} T^2$ where $T$ is the time between pulses.  

There are unique challenges and advantages introduced with LPAI in a warm vapor.  First, the interrogation time $T$ is necessarily limited to tens of microseconds, the time it takes the majority of atoms to transit the centimeter scale Raman laser beam. This significantly diminishes $\Delta\phi$ in comparison to traditional experiments with $T >$ 10 ms.  In principle, this can be compensated with a large signal-to-noise ratio (SNR) due to the availability of a large density of atoms approaching $10^{12}/$cm$^3$, a limit imposed by radiation trapping~\cite{Rosenberry2007}.  On the other hand, a short interrogation time beneficially affords both a high data rate, and a large dynamic range as defined by the acceleration required to cause a $\pi$-radian phase shift.  In this work, we demonstrate an ultra-high data rate of 10 kHz, and an ultra-large dynamic range of 88 \textit{g} where \textit{g} = 9.8 $m/s^2$. Both are orders of magnitude larger than the fastest LPAI accelerometers \cite{McGuinness2012}.  

Second, in a warm vapor, multiple simultaneous atom interferometers can be achieved taking full advantage of the large number of velocity classes available for interrogation (as indicated in \autoref{fig:fig1}).  In our experiment, the maximum number of such interferometers is approximately $N_{AI} \approx 2 |\textbf{k}_{\textrm{eff}}| v_B / \Omega_R =739$ where $v_B$ is the one-dimensional rms velocity of the vapor.  As a proof of this concept, we demonstrate two simultaneous interferometers by observing interference signals from the velocity classes at $\pm 7.8$ m/s $\hat{z}$.    These two interferometers possess equal and opposite sensitivity to acceleration.  Differencing the phase shift of the two interferometers provides common-mode rejection of spurious offset noise and doubled sensitivity to acceleration.

Third, state polarization is crucial for observing clear interference fringes.  Also, spectator atoms colliding with the wall and pumping to $|\text{F} = 2\rangle$ during the measurement can obstruct the detection of interferometer participant atoms. We address this by coating the cell wall with a self-assembled monolayer that preserves the internal quantum state of the atoms throughout collisions \cite{Yi2008}.  The literature is rich with investigations of various spin-preserving coatings \cite{Corsini2013, Guzman2006, Budker2005, Seltzer2010, Balabas2010}, but there are none that report having low vapor pressure.  We use octyldecyltrichlorosilane as our spin anti-relaxation coating due to its high reactivity as compared with other approaches such as octyltrichlorosilane.  We demonstrate that it allows spin polarization exceeding 90\% in $|\text{F} = 1\rangle$ with a spin-relaxation time of 23 ms in  a 2 cm $\times$ 4 cm $\times$ 10 cm rectangular cell.  Furthermore, we find the outgassing to be minor in our interferometer experiments.  

Lastly, the laser frequency requirements are minor when compared with the complexity and agility needed for a laser-cooled atom interferometer \cite{McGuinness2012}.  Our technique requires only 3 $\textit{static}$ laser frequencies for depump, probe and Raman transitions.  These 3 lasers are pulsed sequentially during the interferometer cycle, and require only one optical axis for delivery to the cell.  

\begin{figure}
\includegraphics[width=3.5 in]{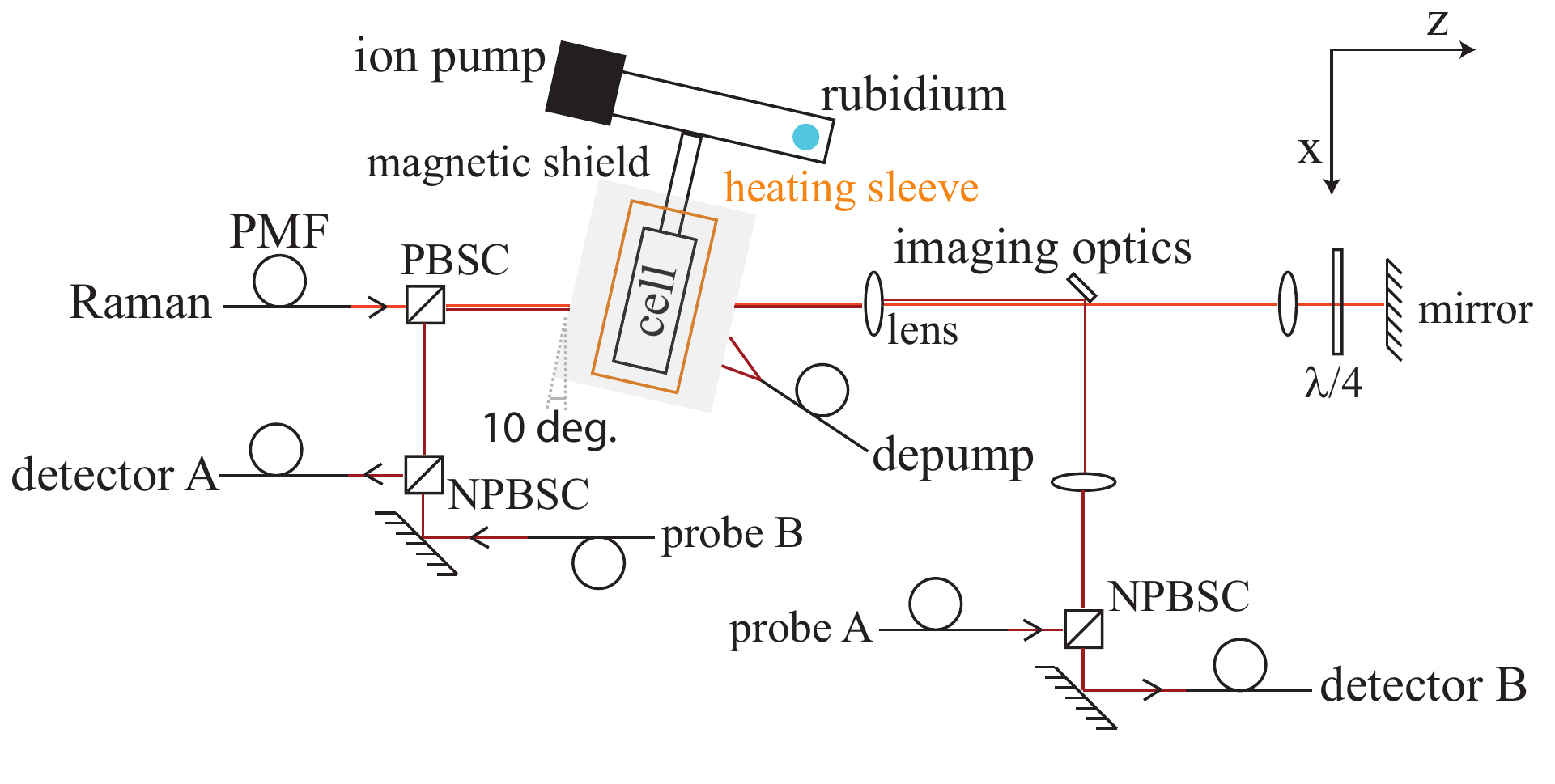}
\caption{\label{fig:fig2} (Color online) Schematic of the experiment  showing the counter-propagating probe lasers, the off-axis application of the depump laser, and the imaging-pickoff scheme for separating the co-linear probe and Raman lasers. The vapor cell sits within a heating sleeve and is surrounded by a magnetic shield.  NPBSC: non-polarizing beam splitter cube, PBSC: polarizing beam splitter cube, PMF: polarization maintaining fiber}
\end{figure}

In further detail, the depump laser, tuned 80 MHz blue of  $|\text{F} = 2\rangle \rightarrow |\text{F'} = 2\rangle$  on the D1 line, optically pumps the vapor into the $|\text{F} = 1\rangle$ manifold.   The two probes (see \autoref{fig:fig2}), both tuned to 10 MHz blue of the $|\text{F} = 2\rangle \rightarrow |\text{F'} = 3\rangle$ resonance on the D2 line, counter-propagate through the cell allowing the simultaneous detection of  two velocity classes.  The Raman laser is detuned 1.208 GHz below the $|\text{F} = 2\rangle \rightarrow |\text{F'} = 2,3\rangle$ crossover transition on the D2 line.    The frequency is stabilized with a beatnote offset lock feeding back to the fiber laser seed.

The experimental setup is detailed in \autoref{fig:fig2}.  The spin preserving coating allows the use of a low depump power to maintain the state preparation of the atoms.  We find it sufficient to expand 10 mW of depump light freely from a fiber into the vapor cell.  The probe beams are collimated to $r_p = 2.8$ mm 1/$e^2$ radius with an intensity of 0.18 mW/cm$^2$ at the cell.  This intensity is well below the saturation value of 1.67 mW/cm$^2$ to favor a linear response of the probe absorption with vapor density.  The probes are coupled to independent detectors using single mode fiber to enhance background signal rejection. The Raman laser is seeded by a telecom fiber laser at 1560 nm that passes through a fiber optic phase modulator operating near the hyperfine splitting, $\nu_{hf} \approx 6.8$ GHz.  This light is amplified in a 30 W fiber amplifier and then doubled to 780 nm in a periodically-poled lithium niobate crystal \cite{Sane2012}.  The Raman beam is collimated to 5.6 mm 1/$e^2$ radius with a peak power of $\approx$ 3 W giving a Rabi frequency of $\Omega_R/2\pi$ = 1.2 MHz.  This beam passes through the cell and a $\lambda/4$ waveplate before retroreflecting forming a linear $\perp$ linear Raman beam polarization to minimize Doppler free Raman excitation.  

\begin{figure}
\includegraphics[width=3.5 in]{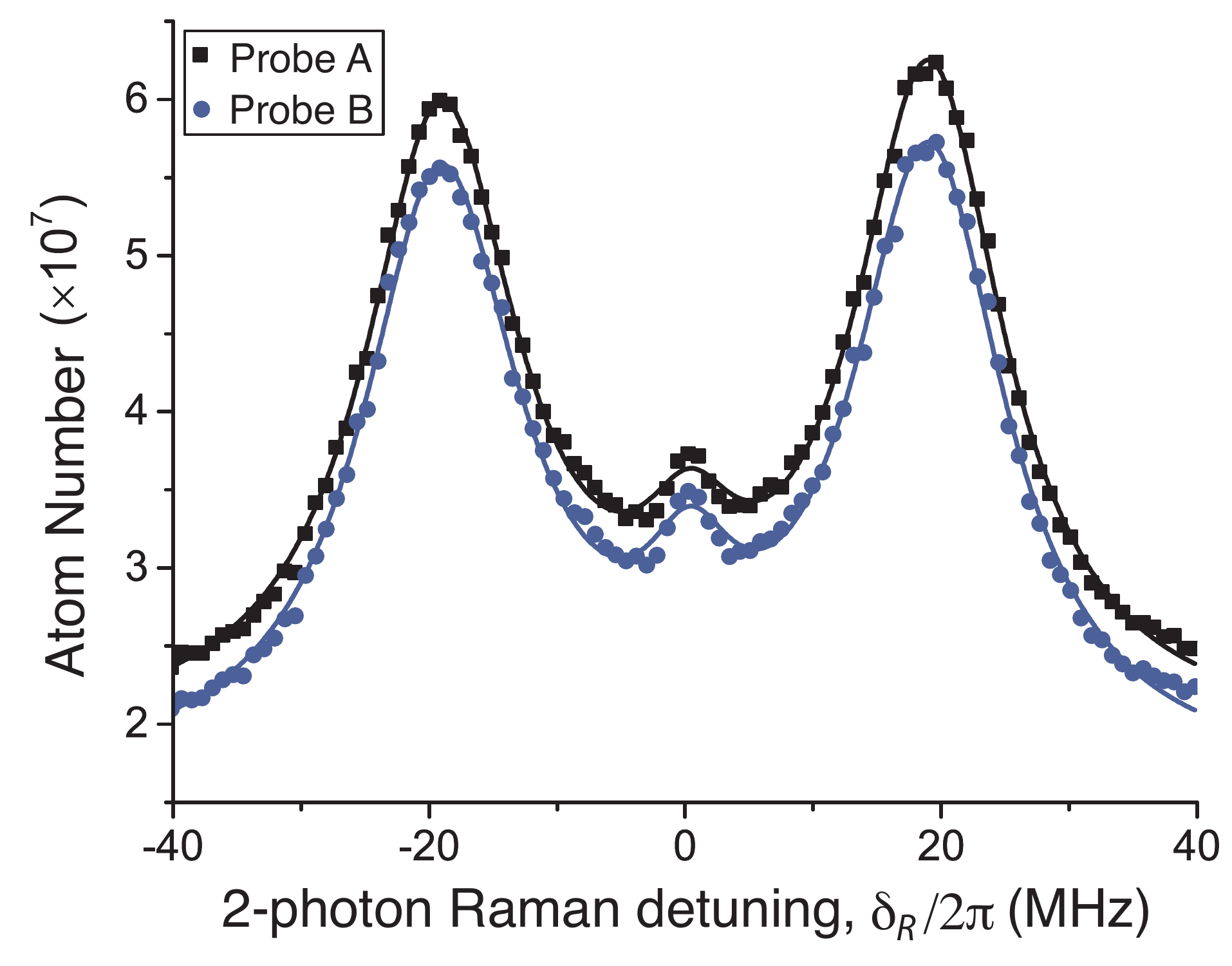}
\caption{\label{fig:fig3} (Color online) Raman detuning frequency scan revealing peaks for $\pm|\textbf{k}_{\textrm{eff}}|\hat{z}$  at $\approx\mp 20 $ MHz for probe A ($\mp|\textbf{k}_{\textrm{eff}}|\hat{z}$ at $\approx\pm 20 $ MHz for probe B) which detects the velocity class centered at $\approx-$7.8 m/s $\hat{z}$ ($\approx+$7.8 m/s $\hat{z}$).  The feature centered at zero is a muted Doppler free resonance due to imperfect optical polarization.  The Raman-excited velocity classes are given by  $\textbf{v}_c  =  \delta_R/\textbf{k}_{\textrm{eff}}$ where $\textbf{k}_{\textrm{eff}}$ is simultaneously along $\pm\hat{z}$ in our experiment. The measured widths of the four peaks are, on average, $14.6\pm1$ MHz, consistent with the expected linewidth from numerical analysis of $\approx(\lambda\gamma|\textbf{k}_{\textrm{eff}}|/2\pi+1.4\Omega_R)/2\pi = 13.8$ MHz for $\Omega_R < (\lambda\gamma|\textbf{k}_{\textrm{eff}}|/2\pi)/8$.}
\end{figure}

We optimize detection of signal atoms by propagating the Raman and probe lasers collinearly to overlap the addressed velocity classes.  Since these lasers are nearly the same wavelength, we use polarization and imaging techniques to combine and separate the two beams (see \autoref{fig:fig2}). The two lasers co-propagate through the cell with a slight angle ($<$ 1 mrad) and overlapped to better than 1 mm.  After transit through the cell, a telescope separates the two beams with a pick-off mirror at the focal plane.  This technique suppresses the Raman power reaching the probe B detector by 7 orders of magnitude which avoids over saturating the detector.  With this setup, in \autoref{fig:fig3} we show signal from a scan of the Raman detuning, $\delta_R$, revealing, for each fixed probe, two Doppler sensitive peaks associated with $\pm |\textbf{k}_{\textrm{eff}}| \hat{z}$.   The peaks are Lorentzian, as the narrow Doppler sensitive resonance ($\Omega_R/2\pi = 1.2$ MHz) scans the broad resonance of the probe transition ($\lambda\gamma|\textbf{k}_{\textrm{eff}}|/(2\pi)^2=12.1$ MHz).

The vapor cell is attached to a vacuum chamber with a rubidium sample and a 5 l/s ion pump.  We find that it is necessary to use the ion pump when the cell is warm to avoid suppression of fringe contrast due to collisions with background gas.  The cell penetrates into a magnetic shield assembly to provide a homogeneous magnetic environment that is zeroed to better than 10 mG with a bias coil assembly interior to the shield.  We increase the  vapor density by heating the cell with a controlled warm air flow to reduce aberrations in the Raman beam wavefront.  The remainder of the chamber is maintained several degrees colder than the cell to avoid buildup of rubidium on the coating which can ruin the efficacy. We empirically find that a vapor density of $n = 4 \times 10^{10} /\text{cm}^3$, corresponding to a temperature of $39~^\circ$C, is an optimal trade off between increased signal and occlusive background from imperfect spin polarization.  

\begin{figure}
\includegraphics[width=3.5 in]{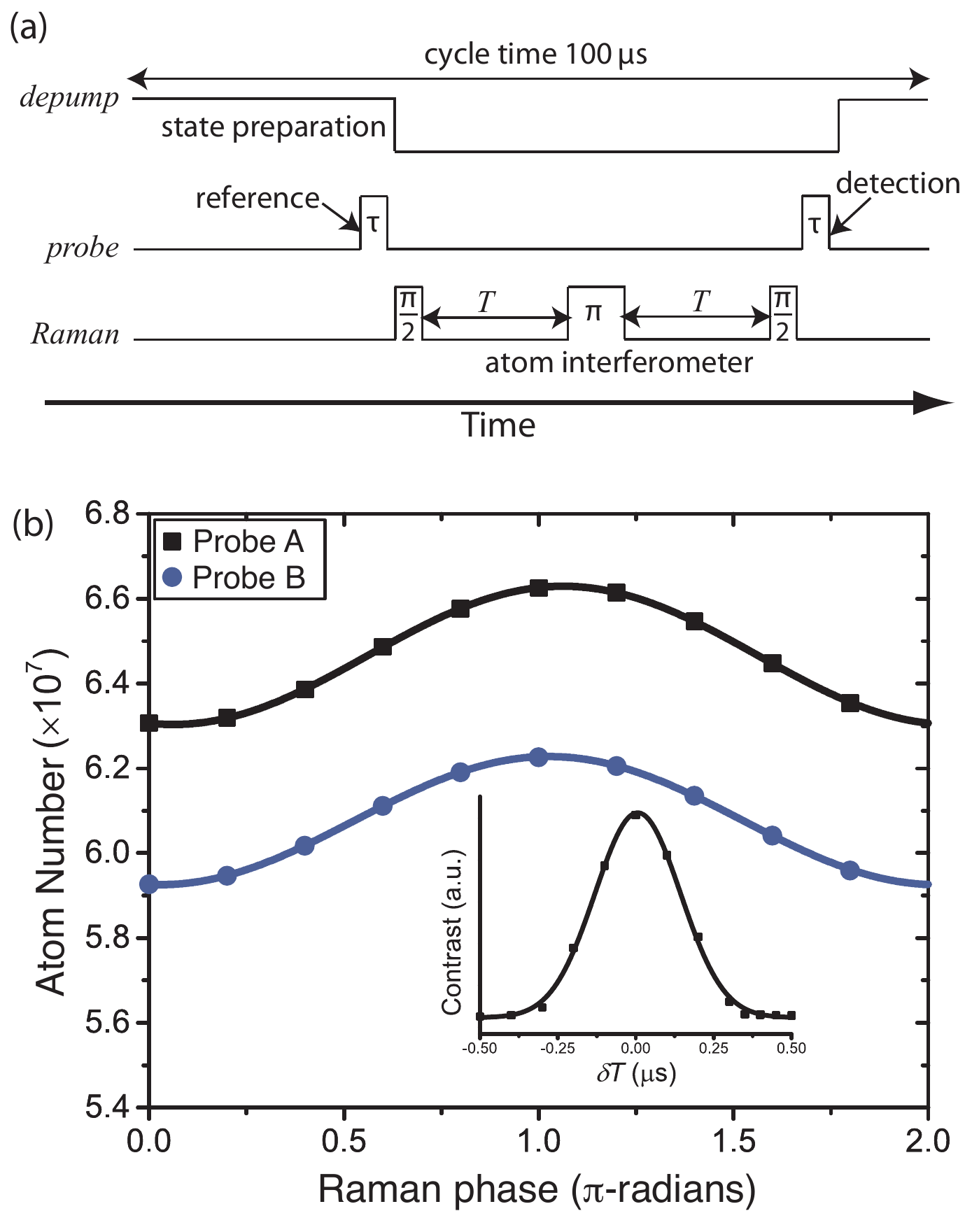}
\caption{\label{fig:fig4} (Color online) a) Timing diagram for the experimental pulse sequence with $T = 15~\mu s$, a repetition rate of 10 kHz and $\tau$ = 3 $\mu$s. b) Sample fringe in $|\text{F}=2\rangle$ resulting from a Raman phase scan of a 10 point fringe at a data rate of 10 kHz.  Each point is an average of 200 shots, and a $\pi$ phase shift corresponds to 88 \textit{g}.  The error bars for each point are below the resolution of the graphic.  Inset: The normalized contrast as a function of the timing asymmetry of the interferometer pulses reveals a velocity width of 0.37(2) m/s rms of the interferometer participant atoms. }
\end{figure}

A timing sequence of the optical pulses in our interferometer experiment is shown in \autoref{fig:fig4}(a). The experimental sequence consists of four steps. First, the atoms are prepared in the $|\text{F}=1\rangle$ manifold with depump light. At the end of the preparation pulse, a background measurement is made to provide an atom number reference. We next apply the Raman pulse sequence realizing the interferometer, followed by a probe pulse to measure the transfer to $|\text{F}=2\rangle$.

The detected atom number is calculated using the ratio of the probe and background reference measurements, and scaled to the signal acquired with an off-resonant probe to account for imperfect spin polarization.  This method immunizes the measurement against slow drifts in detection laser intensity and vapor density.  We demonstrate the warm vapor atom interferometer using a $T = 15~\mu$s interrogation in two simultaneous interferometers as shown in \autoref{fig:fig4}(b).  This fringe is the average of 200 phase scans obtained by switching the optical phase to a scanned phase in between the Raman pulses.  We extract the interferometer phases using a sinusoidal fit to the scanned fringes.  After differencing, we measure a shot-to-shot phase noise of 74 mrad that is equivalent to an acceleration sensitivity of 10 m$g$/$\sqrt{\text{Hz}}$ when accounting for the data rate of 10 kHz.  An Allan standard deviation reveals a signal stability with a minimum of 40 m$g$ at $\approx$ 0.2 s and a long term stability below 100 m$g$ at 10 minutes.  We verify that the observed fringes arise from a Doppler-sensitive process by using an asymmetric pulse sequence ($\pi/2 - (T-\delta T) - \pi - (T+\delta T)-\pi/2$) to measure contrast as a function of wavepacket overlap offset given by $2 \hbar |\textbf{k}_{\textrm{eff}}| \delta T/m$, where $m$ is the mass of $^{87}$Rb \cite{Parazzoli2012}. This reveals an average coherence length of 0.81(3) nm rms, and a velocity width of 0.37(2) m/s rms (see inset of \autoref{fig:fig4}(b)).

We calculate the potential sensitivity of an accelerometer using this technique with both our current conditions as well as ideal conditions.  The fundamental phase uncertainty of each $\Delta\phi$ measurement is given by
\begin{equation}\label{dphi}
    \delta\phi=\frac{\sqrt{N_{\rm detect}}}{N_{\rm i}},
\end{equation}
where $N_{\rm detect}$ is the total number of atoms being detected by the probe laser, and $N_{\rm i}$ is the subset of atoms that are not only detected but also complete the $\pi/2-\pi-\pi/2$ sequence without leaving the detection volume. In steady state, the number of atoms is conserved inside the probe beam volume, which has a path length $l$ and a probe beam radius $r_p$, and hence a volume of $\pi r_p^2l$.  For a selected velocity class $v_z \ll v_B$, the amplitude in this velocity group is approximately $1/\sqrt{2\pi} v_B$.  For Rabi frequencies less than the probe transition linewidth ($\Omega_R/4\pi < \gamma/2\pi$), as is the case here, the velocity class excited from $|\text{F} = 1\rangle$ to $|\text{F} = 2\rangle$ by the interferometer sequence has a width determined by $\Omega_R/|\textbf{k}_{\textrm{eff}}|$.  Furthermore, with a finite optical pumping inefficiency, we expect a background of detected $|\text{F} = 2\rangle$ atoms with a velocity class of $\lambda \gamma/2\pi$. Since the spin relaxation time is much longer than the interrogation time, we find
\begin{equation}\label{Ndetect}
	N_{\rm detect}\approx \left( \frac{\Omega_R}{|\textbf{k}_{\textrm{eff}}|} + \xi \frac{\lambda \gamma}{2\pi}  \right) \frac{n \pi r_p^2l}{\sqrt{2\pi}v_B }.
\end{equation}
Here,  $\xi$ is the optical pumping inefficiency, and $n$ is the vapor density.

Due to the large thermal velocity,  $N_{\rm i}$ will decay with time. The loss is primarily radial since we select a velocity class near zero in the $\textbf{k}_{\textrm{eff}}$ direction.  We approximate the cylindrical volume in rectangular coordinates to enable an analytic solution and model the number of interferometer participant atoms as 
\begin{eqnarray}\label{Ni}
N_{\rm i}  & \approx & \left[ \sqrt{\frac{2}{\pi}} \frac{2T}{\tau} \left( e^{-\frac{1}{2} (\tau/2T)^2} - 1 \right) + \text{Erf} \left( \frac{1}{\sqrt{2}} \frac{\tau}{2T} \right) \right]^{2} \nonumber \\
   && \times  \frac{\Omega_R}{|\textbf{k}_{\textrm{eff}}|} \frac{n \pi r_p^2l}{\sqrt{2\pi}v_B }, 
\end{eqnarray}
where $\tau = \sqrt{\pi}r_{p}/v_B$ is the transverse transit time. Using Eqs.~(\ref{Ndetect}, \ref{Ni}), and our experimental parameters: $\lambda=780$ nm, $v_B=173$ m/s, $\Omega_R/2\pi=1.2$ MHz, $\gamma/2\pi = 6$ MHz, $\tau = 29~\mu$s, $l=4$ cm, $T = 15~\mu s$, $\xi$ = 0.2 \footnote{During high-data rate operation the effective optical pumping efficiency is reduced from 90\% to 80\%}, and the measured $n=4\times10^{10}$ cm$^{-3}$, we find $N_{\rm detect}\approx9.1\times10^7$ and $N_{\rm i}\approx5\times10^6$ approximately matching our measurements. From Eq.~(\ref{dphi}), this leads to an ideal phase noise for a single interferometer of $\delta\phi\approx2.2$ mrad per shot and an acceleration sensitivity of 0.6 m$g/\sqrt{\rm Hz}$ for a 10 kHz data rate.  This is much smaller than our measured value due to excess noise as discussed below.

As an example of potential performance, we consider the limit of perfect optical pumping.  For an optical depth of 1, which for a cell length $l$ = 1 cm implies a vapor density of $5\times10^{10}$ $\text{cm}^{-3}$, one finds an ideal acceleration sensitivity of 82~$\mu g/\sqrt{\text{Hz}}$ for T = $20~\mu s$,  $\pi r_{p}^{2}=1$ $\text{cm}^{2}$, and the same Raman beam power.  The improved performance is due to the minimization of $N_{\rm detect}/N_{\rm i}$ and the increase in cross sectional area.  We note that for a fixed Raman beam power, the predicted sensitivity grows with the cross sectional area without bound.  However, as is the case here, in practice it will be found that the narrowness of the selected velocity class, and thus the burden on optical pumping will limit the achievable performance. In this sense, it would be advantageous to detect on a more narrow optical transition.

The measured phase noise on each interferometer is larger than the predicted value, and can be explained predominantly by Raman intensity noise.  Measured to be 0.3\% per shot, this drives fluctuations in the selected velocity class width resulting in a calculated noise of $\approx$ 160 mrad per shot for a $\pi/2$-pulse, and is consistent with the measured noise of 145 and 138 mrad per shot for interferometers A and B respectively.  Unlike a cold atom interferometer, pulse area noise is not suppressed due to the significant leakage of interferometer participant atoms from the interrogation region.  However, the pulse area noise is common to the two interferometers, and the resultant phase noise cancels by a factor of 2 after differencing.  The imperfect  cancellation likely stems from a mismatch in the two interferometers, and is the subject of further investigation.

In conclusion, we have demonstrated a light pulse atom interferometer in a warm vapor.  We employ the Doppler selectivity of stimulated Raman transitions to filter a narrow velocity class of atoms for the interferometer.  We show that a light-pulse atom interferometer operating in this manner has the advantage of multiple available velocity classes for sourcing simultaneous interferometers.  Under ideal conditions, we forecast a sensitivity below $100~\mu g/\sqrt{\text{Hz}}$.  Our approach functions without the use of laser cooling and trapping, and without an ultra-high vacuum environment making this attractive for simplified measurement systems.

This research was developed with funding from the Defense Advanced Research Projects Agency (DARPA). The views, opinions, and/or findings contained in this article are those of the authors and should not be interpreted as representing the official views or policies of the Department of Defense or the U.S. Government.  Sandia National Laboratories is a multi-program laboratory managed and operated by Sandia Corporation, a wholly owned subsidiary of Lockheed Martin Corporation, for the U.S. Department of Energy's National Nuclear Security Administration under Contract No. DE-AC04-94AL85000.

\bibliography{vapor}

\end{document}